\documentclass[twocolumn,pra,showpacs,superscriptaddress,amssymb,amsmath,amsmath]{revtex4-1}
\usepackage{graphicx}
\usepackage{epstopdf}
\usepackage{bm}
\usepackage{hyperref}
\usepackage{wasysym}

\hypersetup{%
   pdfpagemode=None, 
   pdfstartpage=1,
   pdfstartview=FitH,
   pdfmenubar=true,
   pdftoolbar=true,
   colorlinks = true,
   linkcolor=blue,
   citecolor=blue,
   urlcolor=blue,
   bookmarksopen=false
}

\newcommand{\be}{\begin{equation}}
\newcommand{\ee}{\end{equation}}

\begin{document}

\title{Cold interactions between an Yb$^+$ ion and a Li atom:\\
  Prospects for sympathetic cooling, radiative association, and Feshbach resonances}

\author{Micha\l~Tomza}
\affiliation{Faculty of Chemistry, University of Warsaw, 
  Pasteura 1, 02-093 Warsaw, Poland}
\affiliation{Theoretische Physik, Universit\"at Kassel, 
  Heinrich-Plett-Str. 40, 34132 Kassel, Germany}
\author{Christiane P. Koch}
\affiliation{Theoretische Physik, Universit\"at Kassel, 
  Heinrich-Plett-Str. 40, 34132 Kassel, Germany}
\author{Robert Moszynski}
\affiliation{Faculty of Chemistry, University of Warsaw, 
  Pasteura 1, 02-093 Warsaw, Poland}

\date{\today}

\begin{abstract}
  The electronic structure of the (LiYb)$^+$ molecular ion is
  investigated with two variants of the coupled cluster method
  restricted to single, double, and noniterative or linear triple
  excitations. Potential energy curves for the ground and excited
  states, permanent and transition electric dipole moments, and
  long-range interaction coefficients $C_4$ and $C_6$ are
  reported. The data is subsequently employed in scattering
  calculations and photoassociation studies.
  Feshbach resonances are shown to be measurable, despite
  the ion's micromotion in the Paul trap. Molecular ions can be formed
  in their singlet electronic ground state by one-photon 
  photoassociation and in  triplet states by  
  two-photon  photoassociation; and control of cold atom-ion chemistry
  based on Feshbach resonances should be   feasible. Conditions for
  sympathetic cooling of an Yb$^+$ ion by an ultracold gas of Li atoms
  are found to be  favorable in the temperature range of 
  10$\,$nK to 10$\,$mK; and further improvements using Feshbach
  resonances should be possible. Overall, these results suggest excellent
  prospects for building a   quantum simulator with 
  ultracold Yb$^+$ ions and Li atoms.
\end{abstract}


\maketitle

\section{Introduction}

Trapped ions are a highly controllable system with strong interactions
and thus find many applications, for example in precision
measurements, quantum computing or quantum
sensing~\cite{LeibfriedRMP03,HaffnerPR08,BlattNatPhys12}. 
Currently, a growing number of experiments combines trapped ions with
ultracold
atoms~\cite{ZipkesNature10,ZipkesPRL10,RatschbacherNatPhys12,RatschbacherPRL13,%
SchmidPRL10,GrierPRL,HarterPRL12,RaviNatCommun12,HallPRL11,RellergertPRL11,SullivanPRL12,SmithAPB14},
allowing to study the dynamics of a single ion immersed in an atomic
Bose-Einstein condensate~\cite{ZipkesNature10} or to control chemical
reactions of a single ion and ultracold atoms~\cite{RatschbacherNatPhys12}.
Further prospects include the idea to build a quantum
simulator with a hybrid system of ultracold ions and atoms to emulate
solid-state physics~\cite{GerritsmaPRL12,BissbortPRL13} or to 
form ultracold molecular ions~\cite{HallPRL12}. 

Any experimental proposal based on a hybrid system of ultracold ions
and atoms has to face two issues. The first one is the choice of
trapping method for the ions; the second one is potential  losses due
to chemical reactions between ions and atoms. Although optical
traps for ions have become available~\cite{SchneiderNatPhot10,CormickNJP},
radio-frequency (rf) based Paul traps still constitute the most popular
choice~\cite{PaulRMP90}. Unfortunately, the time-dependent rf potential induces 
micromotion of the ion that limits the minimum temperature that can be
achieved~\cite{CetinaPRL12,ChenPRL14,Krych13}. 

\begin{figure}[tb]
\begin{center}
\includegraphics[width=\columnwidth]{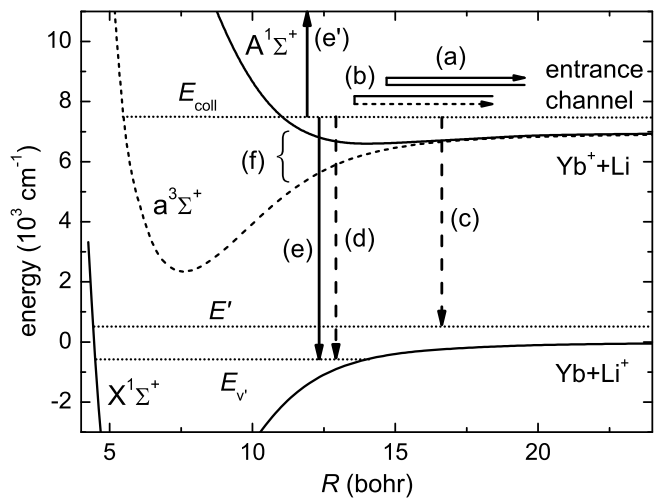}
\end{center}
\caption{Interaction between an Yb$^+$ ion and a Li atom. The
  collision can entail the following processes: elastic scattering
  (a), superelastic (spin-changing) scattering (b), radiative charge
  transfer (c), radiative association (d), photoassociation to ground
  (e) or excited (e') states, and Feshbach resonance based
  magnetoassociation (f).} 
\label{fig:scheme}
\end{figure}

The impact of  micromotion
can be minimized by choosing a large ion to atom mass
ratio~\cite{CetinaPRL12}.  
An Yb$^+$ ion immersed in an ultracold gas of Li atoms is thus a
perfect candidate to construct a hybrid ion-atom experiment. Besides 
the large ion to atom mass ratio to facilitate cooling, 
the relatively simple electronic structure of the (LiYb)$^+$ molecular
ion, with the entrance channel Yb$^+$($^2S$)+Li($^2S$) well separated
from other electronic states, reduces potential nonradiative
losses. Nevertheless, to estimate the rates for radiative loss as
well as the ratio of elastic to inelastic collision cross sections, 
reliable knowledge of the electronic structure is needed. This is
essential both for realizing sympathetic cooling of the ion by the
atoms~\cite{ZipkesNature10,ZipkesPRL10}
and for coupling a crystal of trapped Yb$^+$ ions to a cloud
of ultracold fermionic Li atoms, as proposed for building a quantum
simulator to emulate solid-state physics~\cite{BissbortPRL13}. 
All relevant processes are schematically summarized in Fig.~\ref{fig:scheme}. 

To the best of our knowledge, the electronic structure and collisional
properties of the Yb$^+$ ion interacting with the Li atom have not
been investigated theoretically~\footnote{Note that while our
  manuscript was under review, another study has become
  available~\cite{daSilva15}.}. 
Here we fill this gap and report an 
\textit{ab initio} study of the
interactions
of the Yb$^+$ ion with the Li atom, the system of interest in an 
ongoing experiment~\cite{Gerritsma13}.  
First, we investigate the electronic structure of the (LiYb)$^+$
molecular ion by means of state-of-the-art \textit{ab initio}
techniques. Next, we employ the electronic structure data to
investigate the prospects for sympathetic cooling, spontaneous
radiative association, laser-induced association (photoassociation),
and observation of Feshbach resonances. Special attention is paid to
the formation of molecular ions by means of both spontaneous radiative
association as well as stimulated photoassociation and to the
control of chemical reactivity with both laser and magnetic fields.  

The plan of our paper is as follows. Section~\ref{sec:theory}
describes the theoretical methods used in the electronic structure and
scattering calculations. Section~\ref{sec:results} presents the
results of the \textit{ab initio} calculations in Sec.~\ref{sec:ab_initio},
followed by a discussion of elastic collisions,
radiative charge transfer, radiative association, and  Feshbach
resonances in Secs.~\ref{sec:cooling} to~\ref{sec:Feshbach}. 
A critical assessment of our findings in view of their implication for
experiment, in particular in terms of the prospects
for sympathetic cooling, 
is provided in Sec.~\ref{sec:disc}. 
Sec.~\ref{sec:summary} summarizes our paper.

\section{Computational details}
\label{sec:theory}

\subsection{Electronic structure calculations}

We adopt the computational scheme successfully applied
to the ground and excited states of the SrYb~\cite{TomzaPCCP11} and
Sr$_2$~\cite{SkomorowskiJCP12} molecules as well as the (BaRb)$^+$
molecular ion~\cite{KrychPRA11}. 
The potential energy curve for the $X^1\Sigma^+$ ground electronic
state is obtained with the coupled cluster method restricted to
single, double, and noniterative triple excitations (CCSD(T))~\cite{MusialRMP07}. Potential energy curves for the lowest states
in the $^3\Sigma$ and $^3\Pi$ symmetries are obtained with the
spin-restricted open-shell coupled cluster method restricted to
single, double, and noniterative triple excitations
(RCCSD(T))~\cite{KnowlesJCP99}. Calculations of all other excited
states employ the linear response theory (equation of motion) within the
coupled cluster singles, doubles, and linear triples framework
(LRCC3)~\cite{ChristiansenJCP95,KochJCP97}. The basis set superposition
error is corrected by using the counterpoise correction of Boys 
and Bernardi~\cite{BoysMP70}. The CCSD(T) and RCCSD(T) calculations
were performed with the \textsc{Molpro} suite of codes~\cite{molpro},
while LRCC3 calculations were carried out with the \textsc{Dalton}
program~\cite{dalton}.  
 
The lithium atom is described by the augmented core-valence
correlation consistent polarized valence quadruple-$\zeta$ quality
basis set, aug-cc-pCVQZ. 
For the ytterbium atom, the scalar relativistic effects are accounted 
for by using small-core fully relativistic energy-consistent 
pseudopotential, ECP28MDF~\cite{LimJCP06}, to replace the 
inner-shell electrons, and the associated basis set,
$(15s14p12d11f8g)/[8s8p7d7f5g]$, is employed. 
Due to the less efficient numerical code for the LRCC3 method it was
computationally not feasible to carry out calculations for excited
states correlating all electrons included in the model and using the
large basis set. Therefore in these calculations, the core electrons
are frozen and only the 12 outer-shells electrons are correlated. 
The small-core energy consistent pseudopotential was 
  used also in these calculations, instead of an existing large-core one, to
  avoid potentially different systematic errors for different
  potential energy curves.

The long-range asymptotics of the potentials is computed from
Eqs.~(4)--(8) of Ref. \cite{KrychPRA11}. 
The dynamic polarizability at imaginary frequency of the Yb$^+$ ion and
the Li atom are obtained by using the explicitly connected
representation of the expectation value and polarization propagator
within the coupled cluster
method~\cite{JeziorskiIJQC93,MoszynskiCCCC05} and the best
approximation within the coupled cluster method  
XCCSD4~\cite{KoronaMP06}. The dynamic polarizability at
imaginary frequency of the Li atom is taken from 
Ref.~\cite{DerevienkoADNDT10} and the dynamic
polarizbility of the Yb$^+$ ion is obtained from the sum over state
expression using the transition moments~\cite{SafronovaPRA09}.  
The static quadrupole polarizabilities of the Li and Yb atoms are
taken from accurate atomic calculations reported in
Refs.~\cite{ChenJCP04} and~\cite{PorsevPRA14}, respectively.   

\subsection{Scattering calculations}

In this subsection, we provide a brief summary of all
  equations employed in our scattering
  calculations~\cite{StoofPRB88,MiesJRNIST96,KokooulineJCP99, 
WillnerJCP04,KallushCPL06,ZygelmanPRA88,JulienneJCP78,TellinghuisenJCP84,
MakarovPRA03,CotePRA00}.

The Hamiltonian describing the nuclear motion of the (LiYb)$^+$
molecular ion  reads
\begin{widetext}
  \begin{equation}\label{eq:Ham}
    \hat{H}=-\frac{\hbar^2}{2\mu}\frac{1}{R}\frac{d^2}{dR^2}R+
    \frac{\hat{l}^2}{2\mu R^2}+
    \sum_{S,M_S}V_S(R)|S,M_S\rangle\langle S,M_S|+V^{ss}(R)+V^{so}(R)+
    \hat{H}_{\mathrm{Yb}^+}+\hat{H}_{\mathrm{Li}}\,,
  \end{equation}
\end{widetext}
where $R$ denotes the internuclear distance, $\hat{l}$ the rotational
angular momentum operator, $\mu$  
the reduced mass, and $V_S(R)$ the potential energy curve for the
state with total electronic spin $S$. The relativistic terms $V^{ss}(R)$
and $V^{so}(R)$ stand for, respectively, the spin-dipole-spin-dipole interaction
responsible for the  dipolar relaxation~\cite{StoofPRB88} and the
second-order spin-orbit term~\cite{MiesJRNIST96}. 
The atomic Hamiltonian, $\hat{H}_j$ ($j=$Yb$^+$, Li),
including Zeeman and hyperfine interactions, is given by 
\begin{equation}\label{eq:Ham_at}
\hat{H}_j=\zeta_{j}\hat{i}_{j}\cdot\hat{s}_{j}
  +\left(g_e\mu_{{B}}\hat{s}_{j,z}+g_{j}\mu_{{N}}\hat{i}_{j,z}\right)B_z
\end{equation}
with $\hat{s}_{j}$ and $\hat{i}_{j}$ the electron and
nuclear spin operators, $\zeta_{j}$ denoting the hyperfine coupling
constant, $g_{e/j}$  the electron and 
nuclear $g$ factors, and $\mu_{B/N}$ the Bohr and nuclear
magnetons.  For a fermionic Yb$^+$ ion, Eq.~\eqref{eq:Ham_at}
reduces to the electronic Zeeman term. 
We neglect the unknown second-order spin-orbit coupling $V^{so}(R)$ which has the same form and a similar effect as the spin-spin coupling $V^{ss}(R)$~\cite{StoofPRB88,MiesJRNIST96}.

The bound rovibrational levels are calculated by diagonalizing 
the nuclear Hamiltonian represented on a Fourier grid with adaptive
step size~\cite{KokooulineJCP99,WillnerJCP04,KallushCPL06}. 
The total scattering wave function is constructed in a fully uncoupled
basis set, 
\[
|i_\mathrm{Yb^+},m_{i,\mathrm{Yb^+}}\rangle
|s_\mathrm{Yb^+},m_{s,\mathrm{Yb^+}}\rangle
|i_\mathrm{Li},m_{i,\mathrm{Li}}\rangle
|s_\mathrm{Li},m_{s,\mathrm{Li}}\rangle|l,m_l\rangle\,,
\]
with $m_j$ the projection of angular momentum $j$ on the space-fixed
$Z$ axis, assuming the projection of the total angular momentum
$M_\mathrm{tot}=m_{f,\mathrm{Yb^+}}+m_{f,\mathrm{Li}}+m_l=
m_{i,\mathrm{Yb^+}}+m_{s,\mathrm{Yb^+}}+m_{i,\mathrm{Li}}+m_{s,\mathrm{Li}}+m_l$,
to be conserved. 
The coupled channels equations are solved using a renormalized Numerov
propagator~\cite{JohnsonJCP78} with step-size doubling and about 100
step points per de Broglie wave length. 
The wave function ratio $\Psi_{i+1}/\Psi_{i}$ at the $i$th grid step 
is propagated to 
large interatomic separations, transformed to the diagonal basis, and
the $K$ and $S$ matrices are extracted by 
imposing long-range scattering boundary
conditions in terms of Bessel functions.
 
The rate constant for elastic collisions in the $i$th channel is
given by the diagonal elements of the $S$ matrix summed over partial
waves $l$,  
\begin{equation}
K^{i}_{\rm el}(E) = \frac{\pi\hbar}{\mu k}\sum_{l=0}^\infty(2l+1)\left|1-S_{ii}^l(E)\right|^2\,,
\end{equation}
where $k=\sqrt{2\mu E/\hbar^2}$ with collision energy $E$. 
Similarly, the scattering length is obtained from the $S$ matrix (for $l=0$),
\begin{equation}
  \label{eq:ascatt}
  a=\frac{1}{ik}\frac{1-S_{11}}{1+S_{11}}\,.
\end{equation}

Spontaneous radiative processes are governed by the Einstein coefficients.
For transitions
between two bound rovibrational states $vl$ and $v'l'$, between a scattering
state of energy $E$ and a bound state $v'l'$, and between two scattering states
of energies $E$ and $E'$, they are given by 
  \begin{subequations}\label{eq:Einstein}
    \begin{eqnarray}
      A_{vl,v'l'} &=& \frac{4\alpha^3}{3e^4\hbar^2}
      H_{l} (E_{vl}-E_{v'l'})^3  \Big| \langle\Psi_{vl}|
      d(R)|\Psi_{v'l'}\rangle\Big|^2\,,       \label{eq:Avv}\\
      A_{El,v'l'} &=& \frac{4\alpha^3}{3e^4\hbar^2}
      H_{l} (E-E_{v'l'})^3  \Big| \langle\Psi_{El}|
      d(R)|\Psi_{v'l'}\rangle\Big|^2\,,       \label{eq:AEv}\\
      A_{El,E'l'} &=& \frac{4\alpha^3}{3e^4\hbar^2}
      H_{l} (E-E')^3  \Big| \langle\Psi_{El}|
      d(R)|\Psi_{E'l'}\rangle\Big|^2\,,       \label{eq:AEE}
    \end{eqnarray}    
  \end{subequations}
respectively. 
In Eq.~\eqref{eq:Einstein} the primed and unprimed
quantities pertain to the ground and excited 
state potentials, respectively, $d(R)$ is the transition dipole moment
from the ground to the excited electronic state, $\alpha$
the fine structure constant, and $e$ the electron charge.
The H\"onl-London factor $H_{l}$ is equal to $(l+1)/(2l+1)$ for
the $P$ branch ($l=l'-1$), and to $l/(2l+1)$ for the $R$ branch
($l=l'+1$). In Eq.~\eqref{eq:Einstein}, the scattering states
  $|\Psi_{El}\rangle$  are energy normalized, 
  whereas the wavefunctions of bound levels are normalized to unity, 
  such that the three types of
  Einstein coefficients have different dimensions.

Radiative charge transfer can be described by the following Fermi
golden rule type expression for the rate
constant~\cite{ZygelmanPRA88,JulienneJCP78,TellinghuisenJCP84}, 
\begin{equation}\label{eq:K_ct}
K_\mathrm{CT}(E)=\frac{4\pi^2\hbar^2}{\mu k}\sum_{l=0}^\infty(2l+1)\sum_{l'=l\pm 1}
 \int_0^\infty A_{El,E'l'} d \varepsilon \,,
\end{equation}
where $\varepsilon=E-E'$. Analogously, 
the rate constant for radiative association is given by
\begin{equation}
  K_\mathrm{RA}(E)=\frac{4\pi^2\hbar^2}{\mu k}\sum_{l=0}^\infty(2l+1)
  \sum_{l'=l\pm 1}\sum_{v'}A_{El,v'l'}\,.
  \label{eq:K_ra}
\end{equation}
The total rate constant for radiative losses is the sum of
Eqs.~(\ref{eq:K_ct}) and~(\ref{eq:K_ra}), 
\begin{equation}
  K_{\rm R}(E)=K_\mathrm{CR}(E)+K_\mathrm{RA}(E)\,.
\end{equation}

Stimulated radiative association (photoassociation) becomes possible
by applying a laser field. The rate constant for photoassociation
reads~\cite{SandoMP71,NapolitanoPRL94} 
\begin{equation}
  K_\mathrm{PA}(\omega,E)=\frac{\pi\hbar}{\mu k}\sum_{l}(2l+1)
  \sum_{v'l'}|{S}_{v'l'}(E,l,\omega)|^2\,,
  \label{eq:K_pa}
\end{equation}
with
\begin{equation}
  |{S}_{v'l'}(E,l,\omega)|^2 = \frac{\gamma_{v'l'}^s(E,l)\gamma_{v'l'}^d}
  {(E-\Delta_{v'l'}(\omega))^2+\frac{1}{4}[\gamma_{v'}^s(E,l)+\gamma_{v'l'}^d]^2}\,,
  \label{eq:S} 
\end{equation}
where $\gamma_{v'l'}^{s}(E,l)$ is the stimulated emission rate, and 
$\gamma_{v'l'}^{d}$ the rate for spontaneous
decay. $\Delta_{v'l'}(\omega)$ is the
detuning relative to the position of the bound rovibrational
level $v'l'$, i.e., $\Delta_{v'l'}=E_{v'l'}-\hbar\omega$ with 
$E_{v'l'}$ the binding energy of level $v'l'$. 
The spontaneous emission rates $\gamma_{v'l'}^d$  are obtained from
the Einstein coefficients $A_{v'l',vl}$, 
\begin{equation}
  \gamma_{v'l'}^d=\sum_{vl} A_{v'l',vl}+\sum_{l}
  \int_0^\infty A_{v'l',El}\mathrm{d}\varepsilon\,.
\end{equation}
At low laser intensity $I$, the stimulated emission rate is
given by the Fermi golden rule expression
\begin{equation}
  \gamma_{v'l'}^s(E,l)=4\pi^2\frac{I}{\epsilon_0c}(2l'+1)H_{l'}
  |\langle\Psi_{El}|d(R)|\Psi_{v'l'}\rangle|^2\,.
  \label{eq:gamma_s}  
\end{equation}

Equations~\eqref{eq:K_ct},~\eqref{eq:K_ra}, and~\eqref{eq:K_pa} give
rate constants for a single scattering energy $E$. In practice,  we
have an ensemble of thermally populated states and the rate constants
at a temperature $T$ are obtained by performing a Boltzmann average, 
\begin{equation}
  K(T)=\frac{2}{\sqrt{\pi}(k_BT)^{3/2}}\int_0^\infty K(E)\sqrt{E}\mathrm{e}^{-E/k_BT}dE\,.
\end{equation}
Hereafter, when we speak about the rate constant for a given collision energy we mean $K(E)$, 
  whereas the rate constant for a given temperature implies $K(T)$.

\section{Numerical results and discussion}
\label{sec:results}

\subsection{Potential energy curves, permanent and transition electric
  dipole moments} 
\label{sec:ab_initio}

Before presenting our potential energy curves, and permanent and
transition dipole moments, we first compare the computed atomic
results to the best available experimental data. Our predicted
position of the nonrelativistic $^2P$ state of the Li atom is
14910$\,$cm$^{-1}$, to be compared with the experimental value of
14904$\,$cm$^{-1}$~\cite{nist}. For the $^3P$ state of the Yb atom we
obtain 17635$\,$cm$^{-1}$, in a relatively good agreement with the
experimental value of 18903$\,$cm$^{-1}$~\cite{nist}. 
The predicted ionization potentials (IP) are 43464$\,$cm$^{-1}$  for
Li and 50267$\,$cm$^{-1}$ for Yb, in a good agreement with the
experimental values of 43487$\,$cm$^{-1}$ and
50443$\,$cm$^{-1}$~\cite{nist}, respectively. 
To further assess the quality of the methods, basis sets, and
pseudopotential employed in 
the present work, we have computed the static polarizabilities of the
ground state of the Li and Yb atoms and of the ground state 
of the Li$^+$ and Yb$^+$ ions.
Our calculated polarizability of the Li atom ground state 
amounts to 164.3$\,a_0^3$. The experimental value is
$164.0\pm3.4\,a_0^3$~\cite{PMolofPRA74}, while the best theoretical
result is 164.0$\,a_0^3$~\cite{DerevienkoADNDT10}. Also the static polarizability 
of the Li$^+$ ion, $0.190\,a_0^3$, is in a very good agreement with the
experimental value of $0.188\pm0.002\,a_0^3$~\cite{CookePRA77}. 
Our static polarizabilities of the Yb atom and Yb$^+$ ion,
$143.9\,a_0^3$ and  63.6$\,a_0^3$, also are in a very good agreement
with the most sophisticated atomic calculations, giving
$141\pm6\,a_0^3$~\cite{DzubaJPB10} and
$62.04\,a_0^3$~\cite{SafronovaPRA09}, respectively.   

\begin{figure}[tb]
\begin{center}
\includegraphics[width=\columnwidth]{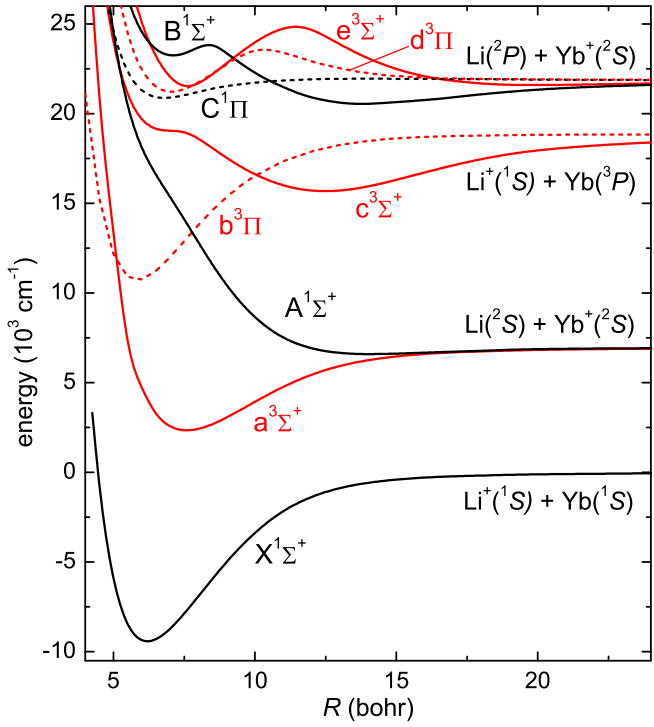}
\end{center}
\caption{(Color online) Non-relativistic potential energy curves of
  the (LiYb)$^+$ molecular ion.} 
\label{fig:curves}
\end{figure}

\begin{figure}[tb]
\begin{center}
\includegraphics[width=\columnwidth]{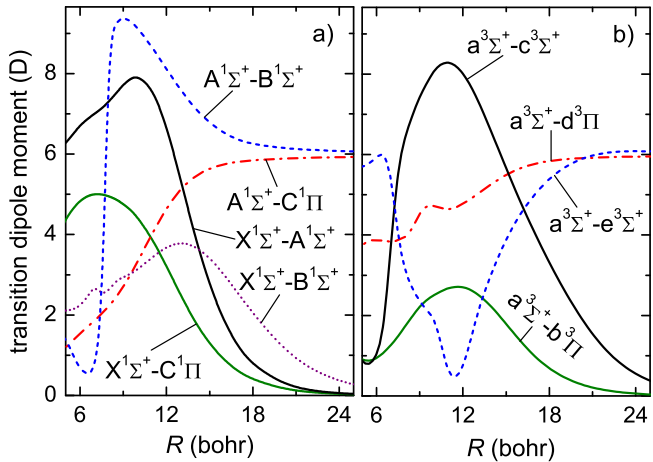}
\end{center}
\caption{(Color online) Transition electric dipole moments between
  singlet (a) and triplet (b) states of the (LiYb)$^+$ molecular ion. 
}
\label{fig:dip}
\end{figure}

\begin{figure}[tb]
\begin{center}
\includegraphics[width=\columnwidth]{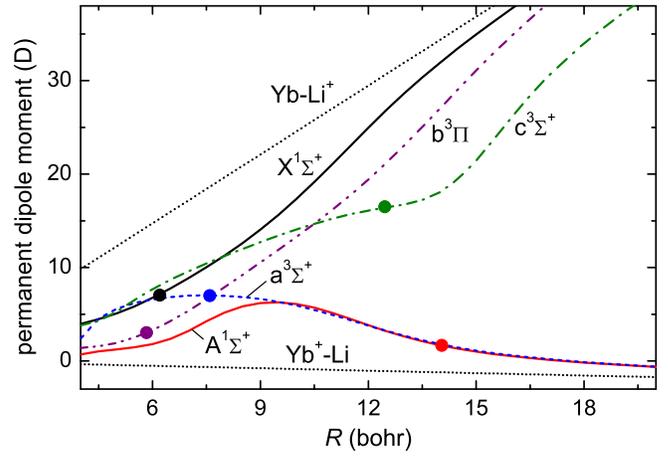}
\end{center}
\caption{(Color online) Permanent electric dipole moments of the most
  relevant singlet and triplet states of the (LiYb)$^+$ molecular
  ion. The $z$-axis is oriented from Yb to Li and the origin is in the
  center of mass. The points indicate the values for the ground
  rovibrational level; and the dotted lines represent the permanent
  dipole moment that the molecular ion would have if the charge was
  completely localized on one of the atoms. 
}
\label{fig:pdip}
\end{figure}

The potential energy curves for the ground and excited states of the 
(LiYb)$^+$ molecular ion are presented in Fig.~\ref{fig:curves}, and
the spectroscopic characteristics are reported in
Table~\ref{tab:spec}. 
The transition and permanent electric dipole moments are plotted in
Fig.~\ref{fig:dip} and Fig.~\ref{fig:pdip}, respectively. 
The leading long-range coefficients for the dispersion and induction
interactions between the Li$^+$ ion and the Yb atom and between the
Yb$^+$ ion and the Li atom, all in the ground electronic state, are
reported in Table~\ref{tab:Cn}.  
\begin{table}[tb]
\caption{Spectroscopic characteristics  of the (LiYb)$^+$ molecular
  ion: equilibrium bond lengths $R_e$, 
  well depths $D_e$, harmonic constants $\omega_0$, and rotational
  constants $B_0$ (for the isotope $^7$Li$^{172}$Yb$^+$). \label{tab:spec}} 
\begin{ruledtabular}
\begin{tabular}{lrrrr}
State & $R_e\,$(bohr) & $D_e\,$(cm$^{-1}$) &  $\omega_0\,$(cm$^{-1}$) & $B_0\,$(cm$^{-1}$) \\
\hline
\multicolumn{5}{c}{Li$^+$(${}^1S$)+Yb(${}^1S$):} \\
$X^1\Sigma^+$ & 6.20 & 9412 & 231 & 0.23    \\
\hline
\multicolumn{5}{c}{Li(${}^1S$)+Yb$^+$(${}^2S$):} \\
$A^1\Sigma^+$ & 14.04 & 358 & 37.1 & 0.045 \\
$a^3\Sigma^+$ & 7.59 & 4609 & 140 & 0.16 \\
\hline
\multicolumn{5}{c}{Li$^+$(${}^1S$)+Yb(${}^3P$):} \\
$b^3\Pi$     & 5.83 & 8130 & 232 & 0.26 \\
$c^3\Sigma^+$& 12.46 & 3177 & 60.9 & 0.057 \\
\hline
\multicolumn{5}{c}{Li(${}^2P$)+Yb$^+$(${}^2S$):} \\
$B^1\Sigma^+$ & 14.02 & 1332 & 50.1 & 0.045 \\
$C^1\Pi$      &  6.71 & 1025 & 138 & 0.20 \\
$e^3\Pi$      &  7.05 & 640  & 170 & 0.18 \\
$d^3\Sigma^+$ &  7.56 & 426  & 218 & 0.16 \\
$d^3\Sigma^+$ & 19.93 & 267  & 24.4 & 0.022
\end{tabular}
\end{ruledtabular}
\end{table}
\begin{table}[tb]
\caption{Induction and dispersion coefficients describing the
  long-range part of the interaction potential between the Li$^+$ ion
  and the Yb atom and between the Yb$^+$ ion and the Li atom, all in
  the ground electronic state.\label{tab:Cn}} 
\begin{ruledtabular}
\begin{tabular}{lrrr}
System &  $C^{\mathrm{ind}}_4\,$($E_ha_0^4$) & $C^{\mathrm{ind}}_6\,$($E_ha_0^6$) &   $C^{\mathrm{disp}}_6\,$($E_ha_0^6$) \\
\hline
Li$^+$+Yb & 72.0 & 1280 & 6.4 \\
Yb$^+$+Li & 82.1 & 711.7  & 711
\end{tabular}
\end{ruledtabular}
\end{table}

The interaction of the ground-state Li$^+$ ion with a ground-state
ytterbium atom results in a single electronic state
- the $X^1\Sigma^+$ electronic ground state of the (LiYb)$^+$ molecular ion, cf. Fig.~\ref{fig:curves}. The
interaction between ion and atom 
is dominated by the induction contribution, resulting in the large
binding energy of 9412$\,$cm$^{-1}$, with the equilibrium distance
equal to 6.2$\,$bohr.  
The interaction of the ground-state Yb$^+$ ion with a ground-state
Li atom, which both are open-shell, results in the two electronic
states $a^3\Sigma^+$ and $A^1\Sigma^+$. The triplet state is strongly
bound with binding energy equal to 4609$\,$cm$^{-1}$, whereas the
singlet potential is weakly bound by only 358$\,$cm$^{-1}$. The large
binding energy of the triplet $a$ state as compared to that of the
singlet $A$ state can be rationalized in the molecular orbitals picture:
The triplet $a$ state is stabilized by an admixture of the antibonding
orbital correlated to the lowest asymptote.
While the order of the LiYb$^+$ molecular states is the same as for
the alkali-metal dimers, unlike in alkali-metal dimers
the singlet ground state and the lowest triplet state are connected to
different asymptotes. Thus the 
singlet $A$ state, which is the first excited singlet state, 
corresponds to excited configurations with higher energies 
than the $a$ state, and therefore, it is weakly bound.  

Estimating the uncertainty of \textit{ab initio}
  calculations is a difficult task, especially for heavy many-electron
  systems. The accuracy depends on the proper treatment of 
  relativistic effects, reproduction of the correlation energy,  and
  convergence in the size of the set of basis functions.  
The method employed here, including the relativistic pseudopotential
and the basis sets used, are
of the same quality as those reproducing the 
potential well depths of the lowest electronic states of the
Rb$_2$~\cite{TomzaMP13} and Sr$_2$~\cite{SkomorowskiJCP12} molecules 
with an error of $3.5\,\%$ or better as compared to experimental results.
Based on this as well as additional convergence analysis for the present
system, we estimate the total uncertainty of the calculated
potential energy curves for the lowest two asymptotes to be about
$5\,\%$ and somewhat larger for the higher excited states. 
The recently available theoretical predictions of the potential well depths
 of the lowest two singlet electronic states of the (LiYb)$^+$ molecular ion~\cite{daSilva15} agree with our results well within 28$\,$cm$^{-1}$ and 144$\,$cm$^{-1}$ for the $X^2\Sigma^+$ and $A^2\Sigma^+$ states, respectively.
The uncertainty of the calculated $C_4$ and $C_6$ coefficients, based
on the accuracy of the used polarizabilities, is at most $5\,\%$. 

Experimental proposals to immerse a ground state ytterbium ion in a
gas of ultracold ground state lithium atoms rely on negligible losses
due to the atom-ion interaction. In this respect, our electronic
structure results, Fig.~\ref{fig:curves}, are promising since 
the corresponding electronic states, $a^3\Sigma^+$ and $A^1\Sigma^+$,  
are well separated from all other electronic
states. This suggests that potential losses due to inelastic charge-transfer
collisions should be smaller than those predicted and observed in
ultracold collisions of a Ba$^+$ ion with Rb atoms where  
strong spin-relaxation was observed due to the spin-orbit coupling of 
both the incoming singlet $A^1\Sigma^+$ and triplet $a^3\Sigma^+$ 
states with the $b^3\Pi$ state at short internuclear 
separation~\cite{KrychPRA11,RatschbacherNatPhys12,RatschbacherPRL13}.

The value of the permanent electric dipole moments calculated with respect to the center of mass and presented in Fig.~\ref{fig:pdip}
increases with internuclear distance. This is typical for
heteronuclear molecular ions. It implies that, in contrast to
neutral molecules, even very weakly bound molecular ions will have a
significant  permanent electric dipole moment. 
The dotted lines in Fig.~\ref{fig:pdip} represent the two limiting cases
where the charge is completely localized on one of the atoms. 
The difference between the dotted lines and the calculated values 
can be understood as the interaction-induced variation of the
permanent dipole moment \cite{HeijmenMP96} or the degree of charge delocalization.
Asymptotically, the permanent dipole moments for all electronic states
have to approach one of the two limiting cases.

\subsection{Elastic collisions}
\label{sec:cooling}

\textit{Ab initio} electronic structure calculations do not provide
sufficiently accurate interaction potentials to predict scattering lengths. 
We therefore calculate the scattering properties for a few isotopic
mixtures with correspondingly different scattering lengths. For our
\textit{ab initio} data, 
the low energy $s$-wave scattering length in the 
$A^1\Sigma^+$ ($a^3\Sigma^+$) state amounts to $-385\,$a$_0$  ($-695\,$a$_0$)
for $^{174}$Yb$^+$+$^6$Li and to $2625\,a_0$  ($982\,a_0$) for
$^{174}$Yb$^+$+$^7$Li.      
For $^{173}$Yb$^+$+$^6$Li and $^{173}$Yb$^+$+$^7$Li the low energy $s$-wave scattering length in the 
$A^1\Sigma^+$ ($a^3\Sigma^+$) state amounts to $-371\,$a$_0$  ($-659\,$a$_0$) and $  2705\,a_0$  ($1042\,a_0$), respectively.
These numbers have to be compared to the characteristic length scale
of the atom-ion interaction~\cite{IdziaszekNJP11,IdziaszekPRA09} which is given by $\sqrt{2\mu C_4/\hbar}$
and amounts to $1218\,$a$_0$ for (LiYb)$^+$. Therefore, the 
$^{174}$Yb$^+$+$^6$Li isotope corresponds to a small background
scattering length and $^{174}$Yb$^+$+$^7$Li to a large one.

\begin{figure}[tb]
\begin{center}
\includegraphics[width=\columnwidth]{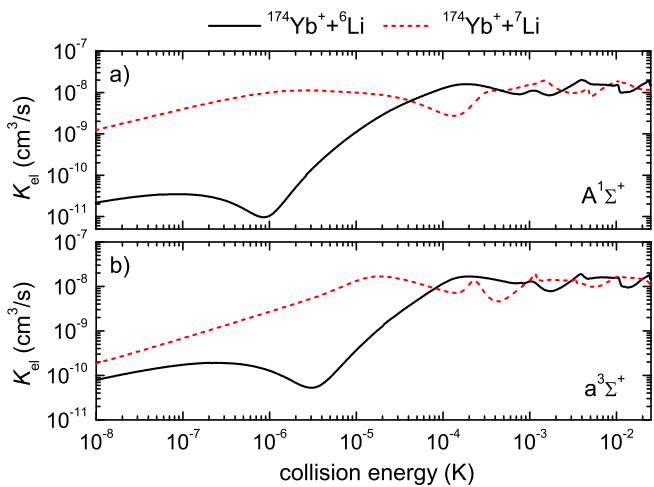}
\end{center}
\caption{Elastic scattering rates for collisions in the $A^1\Sigma^+$
  (a) and   $a^3\Sigma^+$ (b) electronic states. The 
  $^{174}$Yb$^+$+$^6$Li isotope (black solid lines) 
  corresponds to  small background
  scattering length, the $^{174}$Yb$^+$+$^7$Li isotope (red dashed
  lines) to a large one. 
} 
\label{fig:el_cross_sec}
\end{figure}
Elastic collision rate constants are reported in
Fig.~\ref{fig:el_cross_sec}. 
  The height of the centrifugal barrier, for an interaction potential 
  varying as $-C_4/r^4$,  is given by
  $[(l+1)l]^{5/2}/(4C_4^{3/2}\mu^{5/2})$. For example, the barrier
  height amounts to 8.5$\,\mu$K for  $p$-wave collisions. 
For collision energies larger than 10$\,\mu$K, when more partial waves
contribute to the scattering process, the total rate of elastic
collisions is similar for all isotopic mixtures, of the order of
$10^{-8}\,$cm$^3$/s, cf.~Fig.~\ref{fig:el_cross_sec}. 
At lower energies, the elastic collision rates are decreased by two to three
orders of magnitude and depend strongly on scattering length and spin
symmetry. 
Specifically the scattering length and elastic collision rates in the Wigner regime depend on the hyperfine levels of colliding atoms and can be controlled by tuning magnetic Feshbach resonances that we will discuss in Section~\ref{sec:Feshbach}.
Sympathetic cooling is likely to be possible if the ratio 
of the rates for elastic to all inelastic or reactive
collisions is larger than roughly a factor of 100.  
The rates for all inelastic processes are reported in the
following sections. 

\subsection{Radiative charge transfer and radiative association}
\label{sec:ct}

The $A^1\Sigma^+$ electronic state is directly coupled to the ground $X^1\Sigma^+$ electronic
state by the electric
transition dipole moment. This coupling is responsible for potential inelastic losses
due to spontaneous radiative charge transfer (CT), 
\begin{equation}
\mathrm{Yb}^+ + \mathrm{Li} \to \mathrm{Yb} + \mathrm{Li}^+ + h\nu\,, 
\end{equation}
or radiative association (RA), 
\begin{equation}
\mathrm{Yb}^+ + \mathrm{Li} \to \mathrm{LiYb}^+(v,l) + h\nu\,.
\end{equation}
It can also be used for laser-induced molecule formation, i.e., 
photoassociation (PA), 
\begin{equation}
\mathrm{Yb}^+ + \mathrm{Li} +  h\nu' \to \mathrm{LiYb}^+(v',l')
\end{equation}
to the singlet state. The $a^3\Sigma^+$ state is coupled by the
electric transition dipole moment to the  $b^3\Pi$ and
$c^3\Sigma^+$ excited states. These couplings can be used  for 
photoassociation to the triplet states. 

\begin{figure}[t!]
\begin{center}
\includegraphics[width=\columnwidth]{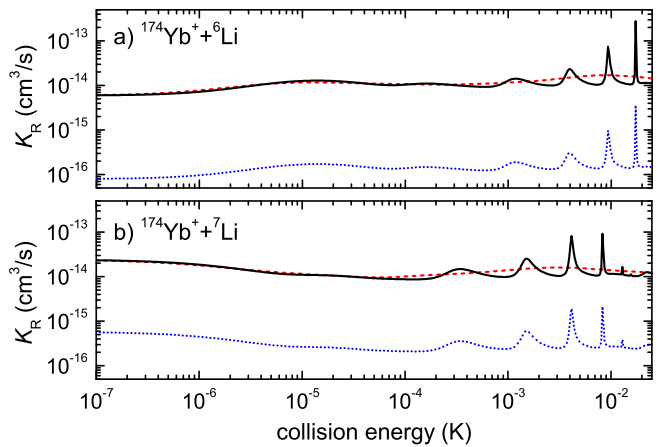}
\end{center}
\caption{(Color online) Rates for radiative
  association (black solid lines) and radiative charge transfer 
  (blue dotted lines) for collisions of an 
  $^{174}$Yb$^+$ ion with  $^6$Li and $^7$Li  atoms. 
  The red dashed envelopes are the thermally averaged rates. 
}
\label{fig:K_r}
\end{figure}
Figure~\ref{fig:K_r} presents the rate constants for radiative
losses in collisions of the $^{174}$Yb$^+$ ion with  $^6$Li and
$^7$Li atoms in the $A^1\Sigma^+$ electronic state (in a field-free case where
all spin orientations are present, the expected rate coefficients are only 25\% of given values). The rates
depend on the presence of resonances in the entrance channel  and follow the Wigner threshold law for small collision energies and the Langevin limit for large collision energies. With rate constants two orders of magnitude larger than for radiative
charge transfer, radiative association represents the main source of
radiative loss that is a common feature for ultracold atom-ion systems~\cite{IdziaszekNJP11,IdziaszekPRA09,SayfutyarovaPRA13,daSilva15,MakarovPRA03}.
At the same time, the rates for 
radiative loss are comparatively small, $10^3$ to $10^5$ times smaller than
the rates for the elastic scattering for temperatures below 10$\,$mK,
cf. Fig.~\ref{fig:el_cross_sec}.
Radiative losses can further be reduced by several orders of magnitude
by applying an external magnetic field which restricts 
the collisional dynamics of the Yb$^+$ ion and the Li atom 
to the high-spin $a^3\Sigma^+$ state. 
Since there is no radiative loss channel for
the triplet state, the only radiative losses for
collisions in magnetic field will originate from an admixture of the 
$A^1\Sigma^+$ state. For an electronic state with a spin-stretched
reference function, the admixture of a singlet state is
given by the long-range dipolar spin-spin interaction and the
spin-orbit coupling with higher excited states.  Both appear in 
second order of perturbation theory, so they are very small~\cite{StoofPRB88,MiesJRNIST96}. For these reasons, radiative losses should not pose a problem for 
sympathetic cooling of Yb$^+$ ions by 
a gas of ultracold Li atoms, especially if an external magnetic
field is applied. 

\begin{figure}[t!]
\begin{center}
\includegraphics[width=\columnwidth]{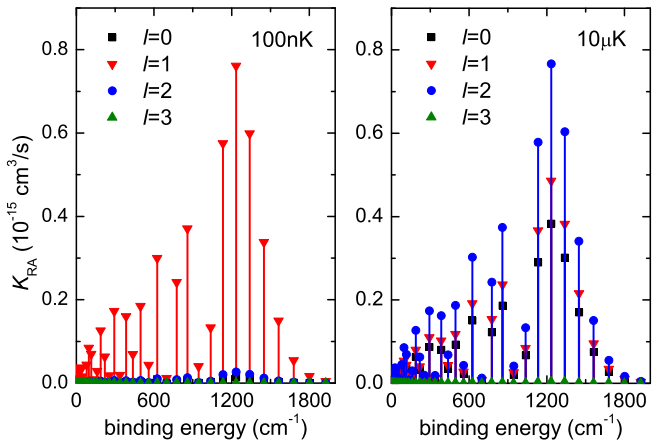}
\end{center}
\caption{(Color online) Radiative association rate   vs binding energy
  of the final vibrational level in the  $X^1\Sigma^+$ ground
  electronic state  for an 
  $^{174}$Yb$^+$ ion colliding with a $^6$Li atom  in the $A^1\Sigma^+$
  state plotted at a temperature corresponding to 100$\,$nK
  (left panel) and to 10$\,\mu$K (right panel). 
}
\label{fig:K_ra_spec}
\end{figure}
Radiative association is further analyzed in 
Fig.~\ref{fig:K_ra_spec} which shows the RA rates to different
rovibrational levels of the ($^6$Li$^{174}$Yb)$^+$ molecular ion at
temperatures of $T=100\,$nK and 10$\,\mu$K. As expected, at 
low temperatures the radiative association
is dominated by the contribution of $s$-wave
collisions, whereas at higher temperatures more partial waves
come into play. The RA spectra for other isotopes have very similar
shapes as those shown in Fig.~\ref{fig:K_ra_spec}
but the amplitudes of the rates vary due to actual value
  of the scattering length and the presence of
resonances in the entrance channel. Interestingly, the largest partial
rates are found for relatively
strongly bound vibrational levels with binding energies of about
1200$\,$cm$^{-1}$. 
The position of the maximum in the radiative association
  spectrum results from the interplay of the overlaps between initial
  scattering and exit vibrational states (related to the structure of electronic states, Fig.~\ref{fig:curves}) and the electric transition dipole moment that decays at large internuclear distances, Fig.~\ref{fig:dip}(a).
Spontaneous radiative association can thus be
used to produce (LiYb)$^+$ molecular ions. In Sections~\ref{sec:photo}
and~\ref{sec:Feshbach} below,  
we will
show that the molecule formation rates can be controlled by means of a
laser field in photoassociation or by a magnetic field modifying
Feshbach resonances. The latter is similar to Feshbach-optimized
photoassociation of neutral molecules~\cite{PellegriniPRL08}. 
But before investigating the control of atom-ion collisions by
external fields, we
examine the sensitivity of our predictions for elastic and
inelastic cross sections on the interaction potential, in particular
on the scattering length. 

\subsection{Sensitivity to the interaction potential}
\label{sec:disc}

Even the best state of the art \textit{ab initio} methods cannot
provide a reliable estimate of the $s$-wave scattering length for many-electron atoms.
Since for (LiYb)$^+$, the $s$-wave scattering length
has not yet been measured, 
it is important to assess the dependency of our results on this
quantity. Most severely, a different value of the scattering length
than that obtained with our data might compromise our conclusions for
sympathetic cooling. Comparison of Figs.~\ref{fig:el_cross_sec}
and~\ref{fig:K_r} suggests sympathetic cooling to be feasible since 
the elastic cross section is three to five orders of magnitude 
larger than the cross
section for all inelastic processes in the relevant temperature range. 
If the true background scattering length is very small, smaller than
that for $^{174}$Yb+$^6$Li (red dashed lines in
Fig.~\ref{fig:el_cross_sec} and upper panel of Fig.~\ref{fig:K_r}), 
the elastic rate constant will be smaller. At the same time the
inelastic losses will also be smaller, due to the decreased
amplitude of the scattering wave function in
the entrance channel. This can be estimated, for example, 
using Eq.~(25) of Ref.~\cite{KrychPRA11}. As a result,  even for
scattering lengths smaller than that of  $^{174}$Yb+$^6$Li in our
calculations,  
the ratio of elastic to inelastic rate constants should still be of
the order of at least 100 for temperatures between 10$\,$nK and
10$\,$mK.
For higher temperatures, when many partial waves
contribute, the overall rate constant is less dependent 
on the $s$-wave scattering length unless a broad shape resonance 
appears. The recently available theoretical results
on the radiative charge transfer and radiative association in the Li+Yb$^+$ system~\cite{daSilva15} reproduce our overall energy dependence and the order of magnitude of the rate constants.

\begin{figure}[tb]
\begin{center}
\includegraphics[width=\columnwidth]{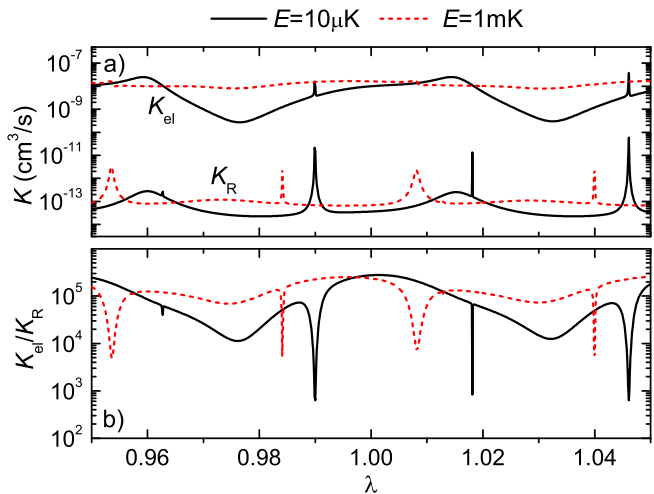}
\end{center}
\caption{(Color online) Sensitivity of the rate constants for elastic scattering $K_\mathrm{el}$ and radiative association and charge transfer $K_\mathrm{R}$ (a) and of the ratio of elastic to inelastic rate constants (b) at collision energies of $10\,\mu$K (black solid lines)  and $1\,$mK (red dashed lines) to a scaling factor $\lambda$ applied to the interaction potential, $V(R)\to\lambda\cdot V(R)$.  
}
\label{fig:ratio}
\end{figure}

For a more detailed assessment of  the sensitivity of our 
results on the uncertainty of the scattering length, we follow the
approach employed in, among others,
Refs.~\cite{ZuchowskiPRA09,GhosalNJP09,SkomorowskiPCCP11}. Specifically, 
we have carried out calculations with the interaction potential scaled by 
a factor $\lambda$, taking values in the range 0.95 to 1.05. This scaling 
roughly corresponds to the estimated error bounds in the calculated 
potential energy curves and changes the number of bound
  states by $\mp1$. The results are shown in Fig.~\ref{fig:ratio} for 
collision energies of 10$\,\mu$K and 1$\,$mK. Inspection of Fig.~\ref{fig:ratio}
reveals a weak dependence of the cross sections on the 
potential scaling factor $\lambda$ to only be interrupted by the presence of sharp 
resonances that occur when bound states of (LiYb)$^+$ cross the incoming 
threshold as a function of $\lambda$. According to
Fig.~\ref{fig:ratio}, these resonances occur in narrow ranges of $\lambda$. The 
probability that the true potential will be such that the ratio of
elastic to inelastic cross sections is seriously affected 
is therefore quite small. Last but not least, the ratio of 
elastic to  inelatic cross section is always much larger than 100.
Within the grid of $\lambda$ employed in the present 
calculations, we thus find strong evidence that  sympathetically cooling 
an Yb$^+$ ion by laser-cooled Li atoms is possible. It is worth
noting that the effect of resonances on the rate constants will be
additionally washed out by  Boltzmann averaging~\cite{HeijmenJCP99}.
The statement predicting feasibility of sympathetic cooling 
can be made even stronger by exploiting the magnetic
field dependence of the scattering
length, discussed in Section~\ref{sec:Feshbach}.

\subsection{Photoassociation }
\label{sec:photo}

Photoassociation refers to the controlled formation of molecules by
applying a laser field~\cite{JulienneRMP06a} and is also a useful tool, e.g. for
spectroscopic investigations.
(LiYb)$^+$ molecular ions can be produced in either singlet or triplet
states:  If the
colliding Yb$^+$ ion and Li atom interact via the $A^1\Sigma^+$
electronic state, photoassociation to the $X^1\Sigma^+$ state
is in principle possible for laser wavelengths between
$611\,$nm and $1438\,$nm.  However, in practice the
  photoassociation window is to a wavelengths between $1050\,$nm and $1438\,$nm
  due to unfavorable Franck-Condon factors. 
If the atom and ion are in the
$a^3\Sigma^+$ state, photoassociation into the manifold of the 
$b^3\Pi$ and $c^3\Sigma^+$ states can be induced with laser wavelengths
between $834\,$nm and $1140\,$nm.
Since the triplet molecules are formed in an electronically excited
state, they are subject to comparatively fast spontaneous decay,
whereas the singlet molecules in the electronic ground state are more
stable. 

\begin{figure}[t!]
\begin{center}
\includegraphics[width=\columnwidth]{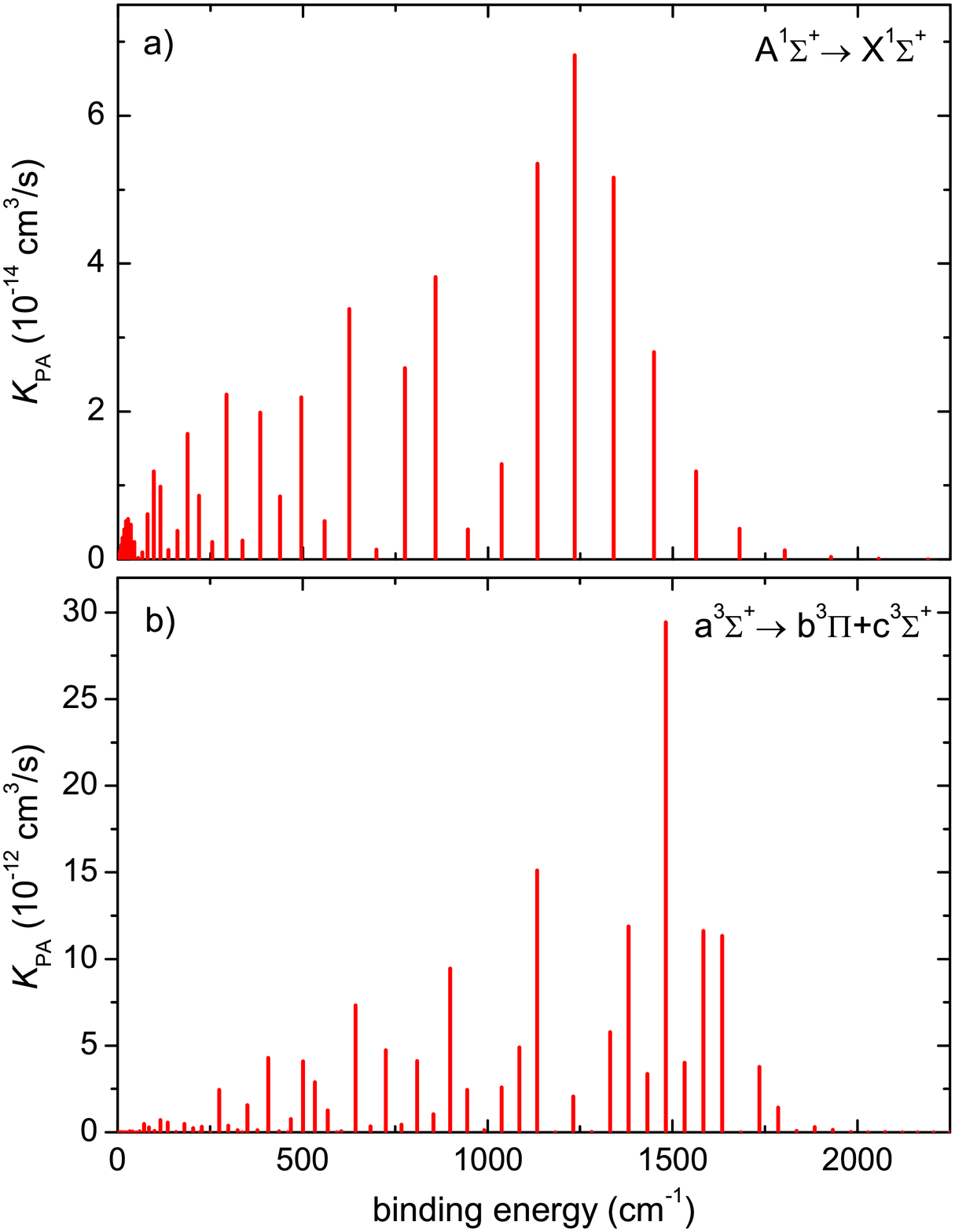}
\end{center}
\caption{(Color online) Photoassociation rates for rovibrational
  levels in the $X^1\Sigma^+$ ground electronic state
  and  $A^1\Sigma^+$  state collisions of 
  $^{174}$Yb$^+$ ions with  $^{6}$Li atoms (upper panel) and 
  for rovibrational levels in the
  $c^3\Sigma^+$ and $b^3\Pi$ states and $a^3\Sigma^+$ state collisions 
  (lower panel) for a laser intensity of $I=100\,$W/cm$^2$ and a
  temperature of 10$\,\mu$K.   
}
\label{fig:PA}
\end{figure}
Photoassociation spectra for $^6$Li$^{174}$Yb$^+$ 
are presented in Fig.~\ref{fig:PA}(a) and~(b) for the singlet and
triplet spin symmetries, respectively.  
In both cases, and similarly to the radiative association discussed above, 
relatively strongly bound rovibrational levels can
be populated, with binding energies of about 1200$\,$cm$^{-1}$ 
(1500$\,$cm$^{-1}$) for the singlet (triplet) symmetry.
The overall shape of the photoassociation spectrum for the singlet symmetry is, as expected, similar to the radiative association spectrum presented in Fig.~\ref{fig:K_ra_spec}.
It is not possible to produce singlet state
(LiYb)$^+$ molecular ions with larger binding energy by using a
one-photon transition since 
the $A^1\Sigma^+$ electronic state is repulsive at 
the equilibrium distance of the $X^1\Sigma^+$ ground
electronic state. To access the $X^1\Sigma^+$ ground vibrational
level, additional Raman transfer would be necessary. 

For triplet state (LiYb)$^+$ molecular ions, 
the largest photoassociation rates  are found for 
ions with a dominant  $c^3\Sigma^+$ state component close to the 
equilibrium internuclear distance. The formation of
triplet molecules  dominated by the $b^3\Pi$ state component
close to the equilibrium internuclear distance (with a binding energy
of about 7000$\,$cm$^{-1}$) is possible. But the rates for these
transitions are three orders of magnitude smaller than the rates for
transitions to the levels dominated by the $c^3\Sigma^+$ state
component.
The lifetime of triplet excited state molecules is
  limited by spontaneous decay.
 Longer lived molecules in the  $a^3\Sigma^+$ state can be
obtained by either spontaneous or stimulated emission. 
These molecules can also be state-electively produced in one step using stimulated Raman adiabatic passage (STIRAP).

The PA spectra for other isotopes have very similar
shapes as those shown in Fig.~\ref{fig:PA}
but the amplitudes of the rates vary due to actual value of the scattering length and
the presence of resonances in the entrance channel.
It is worth noting that the characteristics of the 
spectra for photoassociaton to the lowest singlet and
  triplet states of (LiYb)$^+$ presented in Fig.~\ref{fig:PA},  
 in particular the maxima for relatively deeply bound molecules,
are common for all ion-atom systems when both atom and ion are in the $^2S$ electronic  ground state.  This is in contrast to the photoassociation of alkali metal atoms in the $^2S$ electronic ground state which leads to weakly bound molecules.

The controlled formation of molecular ions is feasible only if the
photoassociation rate  (Fig.~\ref{fig:PA}) is larger than the rates 
for radiative losses (Fig.~\ref{fig:K_ra_spec}). For the
investigated system of Yb$^+$ ion and Li atom, this condition is
met. Additionally, the photoassociation rate can be enhanced by
controlling magnetic Feshbach resonances~\cite{PellegriniPRL08}
which are investigated in the next section.

\subsection{Feshbach resonances}
\label{sec:Feshbach}

\begin{figure}[tb]
\begin{center}
\includegraphics[width=\columnwidth]{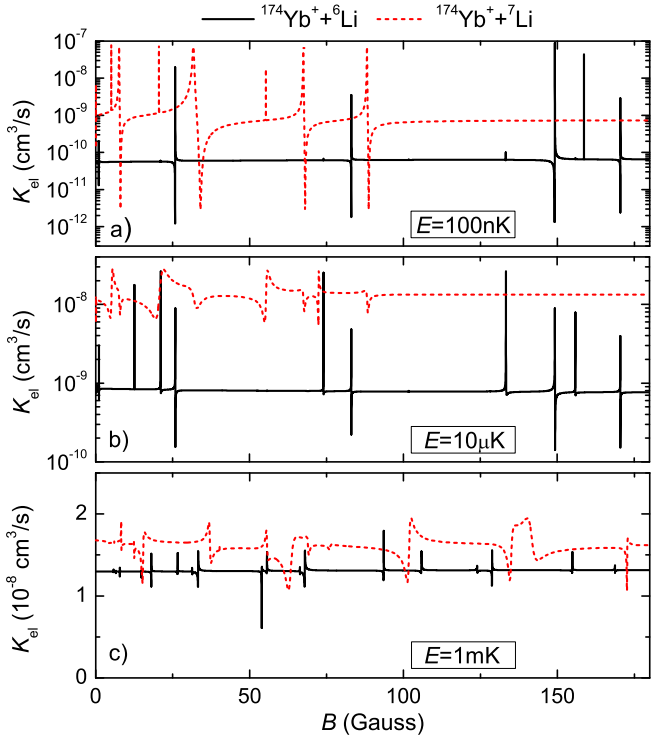}
\end{center}
\caption{(Color online) Elastic scattering rate constant vs magnetic
  field for collisions of  fermionic $^{174}$Yb$^+$
  $(\frac{1}{2},-\frac{1}{2})$ ions with  $^6$Li
  $(\frac{1}{2},\frac{1}{2})$ atoms and $^7$Li $(1,1)$ atoms for collision
  energies corresponding to 100$\,$nK, 10$\,\mu$K, and 1$\,$mK. Note the
  different $y$-axis scales.
}
\label{fig:KelvsB_fer}
\end{figure}

\begin{figure}[tb]
\begin{center}
\includegraphics[width=\columnwidth]{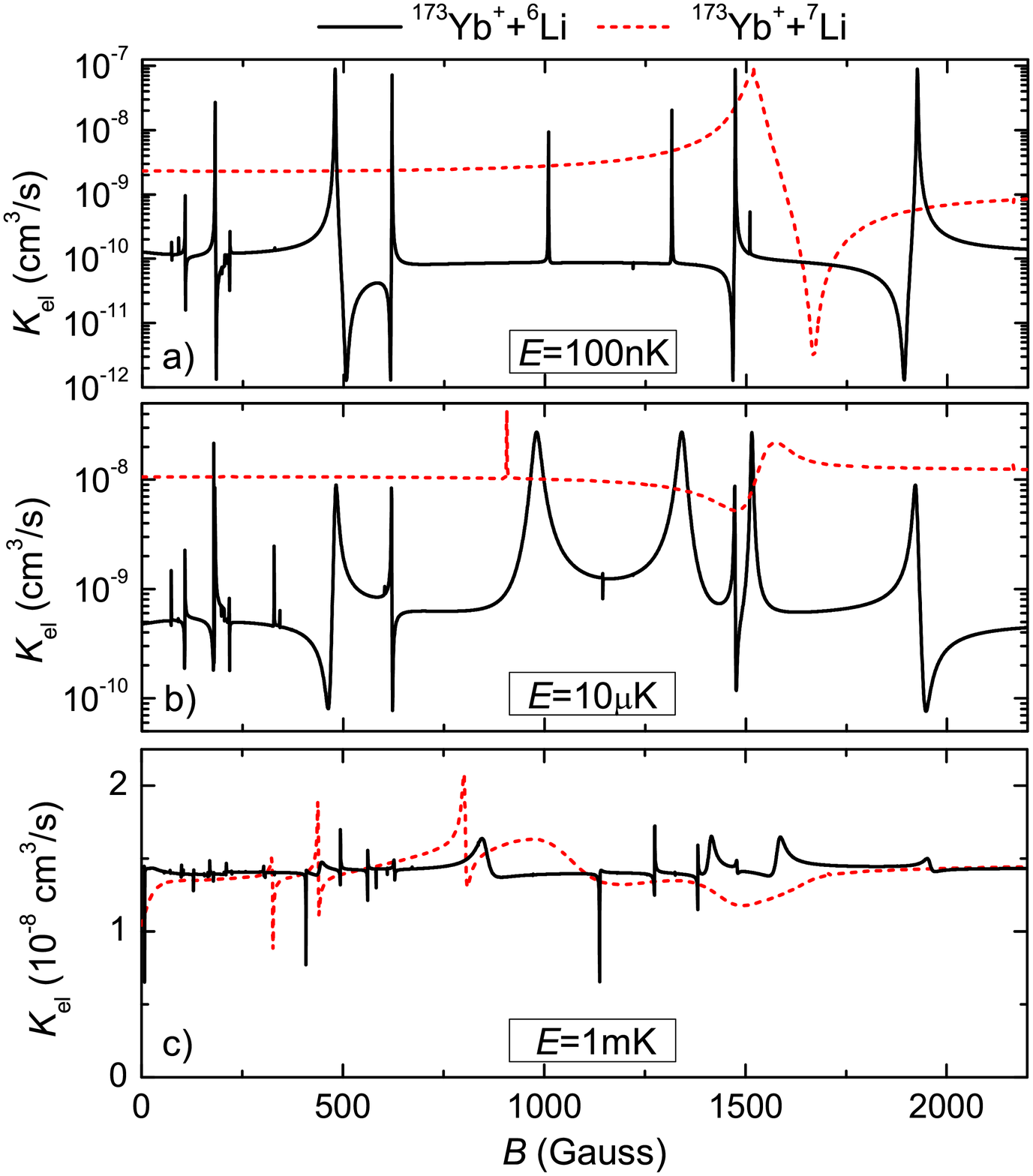}
\end{center}
\caption{(Color online) Elastic scattering rate vs magnetic field for
  collisions of bosonic $^{173}$Yb$^+$ $(3,-3)$ ions with $^6$Li
  $(\frac{1}{2},\frac{1}{2})$ atoms and  $^7$Li $(1,-1)$ atoms for collision
  energies corresponding to 100$\,$nK, 10$\,\mu$K, and 1$\,$mK. 
}
\label{fig:KelvsB_bos}
\end{figure}

Two types of magnetically tunable Feshbach resonances between the
Yb$^+$ ion and the Li atom exist, depending on the structure of the Yb$^+$
ion: 
(i) Fermionic Yb$^+$ ions do not have nuclear spin and
therefore they do not possess any hyperfine structure. Feshbach
resonances between fermionic Yb$^+$ ions and Li atoms result
from the interaction of the hyperfine structure of the Li atom with
the electronic spin of the Yb$^+$ ion~\cite{IdziaszekNJP11,IdziaszekPRA09}. 
(ii) Bosonic Yb$^+$ ions ($^{171}$Yb$^+$ and $^{173}$Yb$^+$) have
a non-zero nuclear spin. As a consequence, they possess a hyperfine
manifold, such that Feshbach resonances between bosonic Yb$^+$ ions
and Li atoms are of the same nature as those between two alkali metal
atoms~\cite{JulienneRMP10}.
The density of the Feshbach resonances at small magnetic fields in the
former case is larger compared to the latter case,
cf.~Fig.~\ref{fig:KelvsB_fer} and Fig.~\ref{fig:KelvsB_bos}.
 Note the different $x$-axis scales for the magnetic 
field in Fig.~\ref{fig:KelvsB_fer} and Fig.~\ref{fig:KelvsB_bos}.
The impact of the Lorentz force and Landau quantization effects~\cite{SimoniJPB11} is neglected in the present study.    

Figure~\ref{fig:KelvsB_fer} shows the rates for the elastic scattering
of the fermionic $^{174}$Yb$^+$ ion in the
$(f,m_f)=(\frac{1}{2},-\frac{1}{2})$ state with fermionic $^6$Li
$(\frac{1}{2},\frac{1}{2})$ and bosonic $^7$Li $(1,1)$ atoms for 
collision energies corresponding to 100$\,$nK, 10$\,\mu$K, and 1$\,$mK
as a function of the external magnetic field.  
Contributions from the lowest three, seven, and fifteen partial waves
are included, respectively, at these three energies. 
The density of Feshbach resonances for the fermionic $^6$Li atom is
larger compared to bosonic $^7$Li because the fermionic
atom has a smaller hyperfine splitting than the bosonic one,
228.2$\,$MHz versus~803.5$\,$MHz. The scattering at temperatures of
100$\,$nK is dominated by  $s$-wave collisions. 
The large Feshbach resonances visible in Fig.~\ref{fig:KelvsB_fer}(a) are almost
pure $s$-wave resonances, whereas the very narrow resonances 
are of $p$-wave character. 
The widths of these resonances are relatively small,
  mostly below 1$\,$Gauss.
At a temperature of 10$\,\mu$K, $s$-wave and $p$-wave collisions
contribute equally, cf. Fig.~\ref{fig:KelvsB_fer}(b), and at
temperatures above 100$\,\mu$K, more partial waves start to contribute
to the elastic scattering. At a temperature of 1$\,$mK, as presented
in Fig.~\ref{fig:KelvsB_fer}(c), the elastic scattering is dominated
by higher partial waves ($l=3$ and $l=4$) which wash out the resonance
structure of the lower partial waves. Thus the Feshbach resonances 
visible  in Fig.~\ref{fig:KelvsB_fer}(c) are
narrower and have much smaller amplitude as compared to those at lower
temperatures.

 
Elastic scattering rates for bosonic $^{173}$Yb$^+$ $(3,-3)$ ions and
fermionic $^6$Li $(\frac{1}{2},\frac{1}{2})$ atoms, respectively, bosonic $^7$Li
$(1,-1)$ atoms as a function of the external magnetic field are shown
in Fig.~\ref{fig:KelvsB_bos} for collision energies of
100$\,$nK, 10$\,\mu$K, and 1$\,$mK. Similarly to
the case of the fermionic $^{174}$Yb$^+$ ion,
cf. Fig.~\ref{fig:KelvsB_fer}, scattering at a
temperature of 100$\,$nK is dominated by $s$-wave collisions with
a few narrow $p$-wave resonances visible. 
 The widths of these resonances are of the order of a few to several Gauss.
At 10$\,\mu$K, $s$-wave and
$p$-wave collisions contribute equally and at temperatures above 
1$\,$mK many partial waves contribute to the rate of the elastic
scattering.  The structure and strength of the Feshbach resonances at
temperatures above 1$\,$mK is washed out by dominant contributions of
higher partial waves to the total elastic scattering rate. 

\begin{figure}[tb]
\begin{center}
\includegraphics[width=\columnwidth]{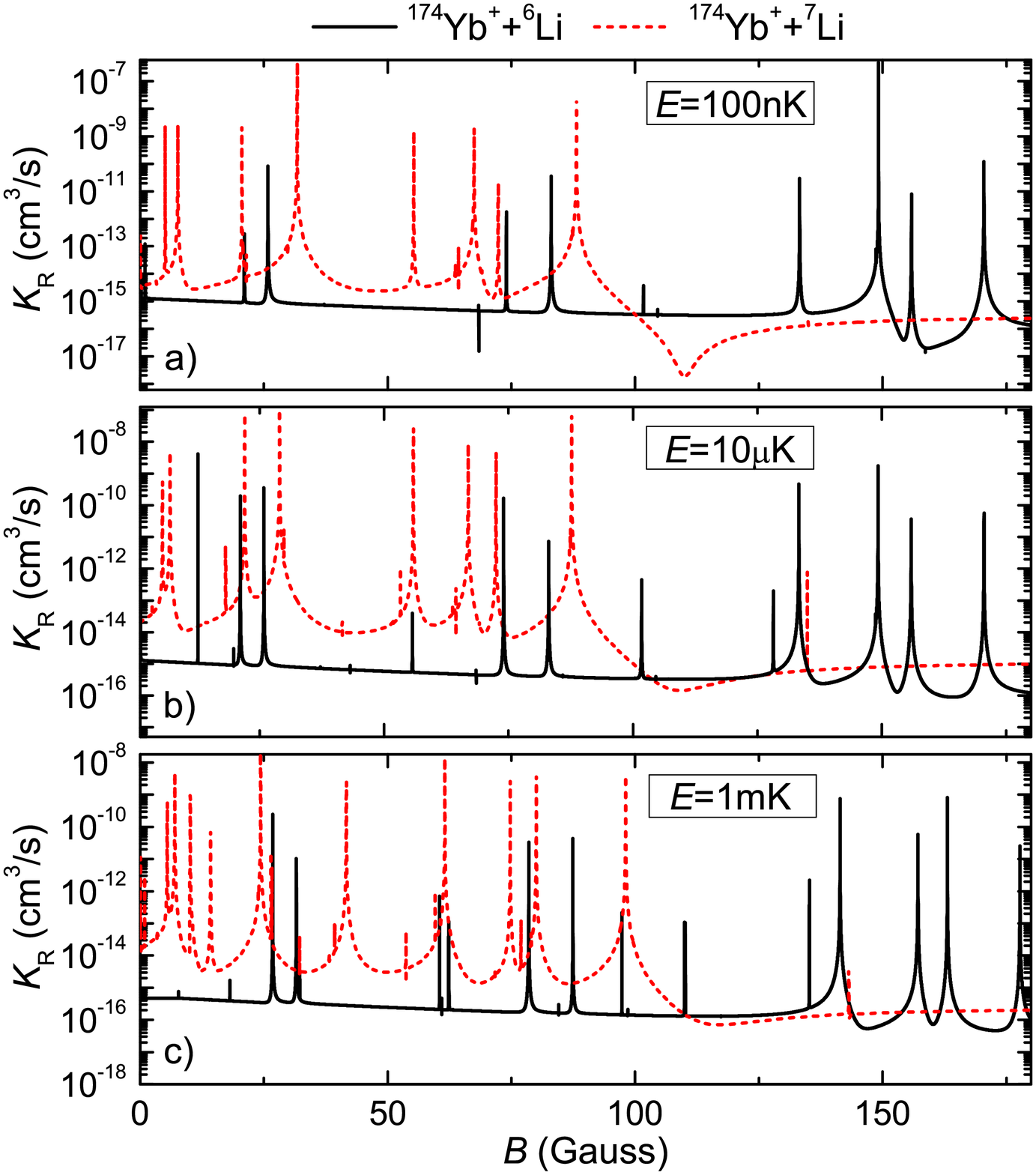}
\end{center}
\caption{(Color online) Radiative loss rate vs magnetic
  field for collisions between  $^{174}$Yb$^+$
  $(\frac{1}{2},-\frac{1}{2})$ ions and $^6$Li
  $(\frac{1}{2},\frac{1}{2})$, respectively, $^7$Li $(1,1)$ atoms for
  collision 
  energies corresponding to 100$\,$nK, 10$\,\mu$K, and 1$\,$mK. 
}
\label{fig:KinelvsB_fer}
\end{figure}
\begin{figure}[tb]
\begin{center}
\includegraphics[width=\columnwidth]{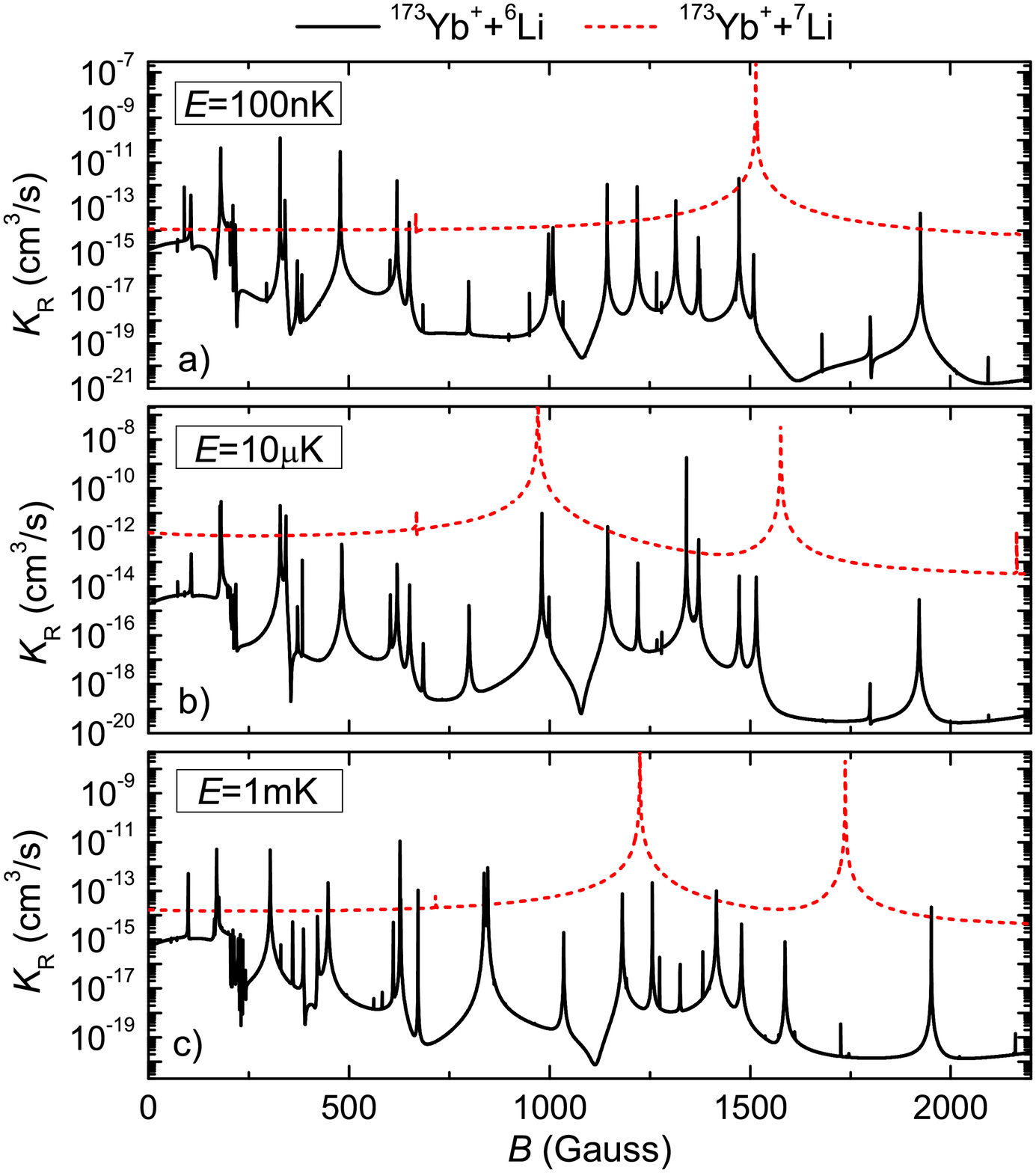}
\end{center}
\caption{(Color online) Radiative loss rate vs magnetic
  field for collisions between  $^{173}$Yb$^+$ $(3,-3)$ ions and 
  $^6$Li $(\frac{1}{2},\frac{1}{2})$, respectively, $^7$Li $(1,-1)$
  atoms for 
  collision energies corresponding to 100$\,$nK, 10$\,\mu$K, and 1$\,$mK. 
}
\label{fig:KinelvsB_bos}
\end{figure}
Finally, Figs.~\ref{fig:KinelvsB_fer} and~\ref{fig:KinelvsB_bos}
present the rates for radiative losses, that is, radiative
association and radiative charge transfer, as a function of magnetic
field for collisions of the same species at the same temperatures 
as in Figs.~\ref{fig:KelvsB_fer}
and~\ref{fig:KelvsB_bos}. In contrast to the elastic scattering rates,
Feshbach and shape resonances are clearly visible for
all investigated temperatures. Far from resonances the typical rates for
elastic scattering are, as expected, at least three orders of magnitude larger than
the rates for  radiative losses. 

Comparing Figs.~\ref{fig:KelvsB_fer} and~\ref{fig:KelvsB_bos} to
Figs.~\ref{fig:KinelvsB_fer} and~\ref{fig:KinelvsB_bos}, a
different temperature dependence of the rates for elastic
scattering and radiative losses is observed. This 
can be understood by analyzing the
mechanisms of the two processes: The elastic rates are dominated by
reflection from the long range potential. For higher partial waves
with a large centrifugal barrier which dominate at high collision
energies, the appearance of Feshbach resonances is strongly
suppressed. By contrast, radiative losses occur at short internuclear
distance only. For high partial waves, this requires tunneling through
the centrifugal barrier. The probability of tunneling decays
exponentially with the height of the centrifugal barrier. Therefore, at
higher temperatures, higher partial waves contribute less to the
radiative losses than to elastic scattering. 

For other isotopes, both fermionic and bosonic ones, 
the characteristics of Feshbach resonances 
such as the density of resonances, the temperature dependence of elastic and inelastic cross
  sections, and the feasibility of controlling the ion-atom interaction with an external
  magnetic field are similar to those discussed above. In particular, they 
depend on the choice of the entrance channel characterized by the
hyperfine states of atom and ion and the scattering length.
Different isotopes of Yb can be
used to control the structure of Feshbach resonances due to the
different masses. However, this control is somewhat limited since, due to the much smaller mass of
  the Li atom, the scattering length depends less on the mass of the Yb atom as compared to mass-balanced mixtures. 

The magnetic field dependence of the scattering length can be exploited to ensure the condition needed for sympathetic cooling.
Figures~\ref{fig:KinelvsB_fer} and~\ref{fig:KinelvsB_bos}
show minima in the radiative loss constants~\cite{HutsonPRL09}, for example close to 
160$\,$G in Fig.~\ref{fig:KinelvsB_fer} (top and middle panel). Tuning
the magnetic field to such a value will enhance the ratio of elastic to
inelastic rate constants by a factor of 1000 and should yield very
efficient sympathetic cooling. Another option consists in tuning the
magnetic field close to one of the broader Feshbach resonances,
observed in Figs.~\ref{fig:KelvsB_fer} and~\ref{fig:KelvsB_bos}, to
increase the elastic rate constant while keeping inelastic losses at
bay. 
The exact positions of the Feshbach resonances will depend on the true
value of the scattering length. However, the characteristics of the
presented results, such as the density of resonances,
are general and depend essentially  
on the hyperfine structure and nature of colliding ion and atom. 
We therefore expect our predictions to be valid, independent of the specific value of the background scattering length. 

The magnetic field dependence can also be employed to enhance the
inelastic rate constants, in case one is  interested in the observation of
cold chemical reactions. If the field-free scattering length is
essentially zero, this will decrease the inelastic rate
constants, as explained above. The remedy then is to tune the magnetic
field to a value where a maximum in $K_R$ is observed, cf.
Figs.~\ref{fig:KinelvsB_fer}
and~\ref{fig:KinelvsB_bos}. Alternatively, one could also make use of 
minima in the elastic rate constants, cf. Fig.~\ref{fig:KelvsB_fer},
which will decrease the ratio of the rate constants by up to a factor
of 1000. The latter is not applicable at higher temperatures
($>10\,\mu$K) where the Feshbach resonances are 
less pronounced in the elastic rate constant.

\section{Summary and conclusions}
\label{sec:summary}

We have carried out state of the art \textit{ab initio} calculations
to determine the electronic structure of the (LiYb)$^+$ molecular
ion. Potential energy curves, transition and permanent electric dipole
moments, and long range coefficients were calculated. Good agreement
between our computed atomic results and the available experimental data
was obtained. 

The entrance channels for collisions between the Yb$^+$(${}^2S$) ion
and the Li($^2S$) atom are found to be well separated from all
other electronic states, suggesting comparatively small losses due to
reactive collisions. 
Significant permanent electric dipole moments are obtained, 
even for very weakly bound states of the molecular ion. This 
opens the way for various control schemes. For example, a non-resonant
laser field will lead to a strong Stark effect that can be used to
control energy level shifts state-selectively. This should enable
state-selective detection of the products of the atom-ion chemical
reactions.  

We have subsequently employed the electronic structure data to
calculate the rate constants for  elastic scattering as well as 
radiative charge transfer and radiative association with and without
magnetic field. The relevant temperature scale is set by 
micromotion in the Paul trap. 
For Li+Yb$^+$, it is less severe than for other
species due to the favorable ion to atom mass ratio, limiting the trap
temperature to about
10$\,\mu$K~\cite{CetinaPRL12,ChenPRL14,Krych13}. At this temperature,  
we find clear signatures of Feshbach resonances. This
suggests that it should be possible to observe these resonances 
and use them to control the atom-ion interactions, facilitating
sympathetic cooling or molecule formation. This encouraging 
finding for Yb$^+$ in Paul trap, immersed
in a gas of ultracold Li atoms, is in contrast to many other 
atom-ion species~\cite{IdziaszekNJP11,IdziaszekPRA09}.

Temperatures lower than 10$\,$mK require sympathetic cooling. 
Our results predict a sufficiently large ratio of elastic to inelastic
scattering cross sections, and thus feasibility of sympathetic
cooling, in the temperature range between 10$\,$mK and 10$\,\mu$K for
both small and large values of the scattering length. Prospects for
sympathetic cooling all the way down to 10$\,$nK for the 
  ion trapped in an optical dipole trap 
 also look good. However, at
these very low temperatures, resonances may have a detrimental effect,
and our predictions come with some uncertainty due
to the unknown exact value of the scattering length. 
We have critically assessed the 
validity of our results by scaling the interaction potential, 
assuming an accuracy of the {\em ab initio} data of $\pm 5\%$. 
Except for extremely narrow ranges of the scaling factor, we find the
ratio of elastic to inelastic scattering cross sections to be larger
than 100. The probability that the true interaction potential
corresponds to an unfavorable value of the scaling factor is
very small, and such an unfortunate case could be remedied by
changing the isotope. 
We are thus confident to predict  feasibility of 
sympathetic cooling of an Yb$^+$ ion by Li atoms down 
to temperatures of about 10$\,$nK. This implies excellent 
prospects for building a quantum simulator using Yb$^+$ ions 
immersed in a gas of ultracold Li atoms, bringing quantum 
simulation of solid-state physics with AMO experiments a 
step closer to  reality~\cite{GerritsmaPRL12,BissbortPRL13}. 

Moreover, since radiative association is found to dominate  radiative
inelastic processes, (LiYb)$^+$ is also a good candidate for observing
ultracold atom-ion chemical reactions. \textit{Controlled} formation of
molecular ions becomes possible by applying a laser field, accessing
vibrational levels in the ground electronic state as well as
electronically excited triplet states. 
Photoassociation can utilize one-photon transitions to levels with
moderate binding energies. The formation of molecular ions in their
absolute ground state will require additional Raman transitions. 

Our calculated photoassociation spectra
are also important for optical trapping since they indicate that
photoassociation losses are likely to occur at wavelengths close to
1064$\,$nm. Since photoassociation lines are typically quite narrow,
it should be possible to avoid these losses by proper tuning of the
trapping laser. Our calculations indicate in which regions transitions
are to be expected but would need to be corroborated by spectroscopy
for quantitative predictions. 

Molecular ions offer a particularly interesting perspective for 
photoassociation and related spectroscopies, in that state-selective 
detection of a
single molecular ion in the trap should be possible. Such a detection
scheme can be based on a change of the 
trapping frequency due to the change of the ion mass when the bond is
formed or broken. State
selectivity could be achieved by exploiting the differences in the
Stark shifts of different rovibrational levels which modify 
the spectrum of the trapped ion. One can thus envision a  series of
photoassociation and dissociation experiments, detected by changes in
the fluorescence wavelength. Similarly, one can photo-induce, and
potentially control, ion-neutral chemical reactions by laser
excitation of Li into the $^2P$ state. Due to the favorable shape of
the corresponding $^1\Pi$ state, this might provide an alternative and
more direct route for producing molecular ions in their absolute
ground state than photoassociation into the $X^1\Sigma^+$ ground electronic
state. 

To conclude, Li+Yb$^+$ represents an extremely promising example of 
hybrid atom-ion systems. Good prospects for  sympathetic cooling
should allow for its application in quantum simulation in the not
too distant future~\cite{GerritsmaPRL12,BissbortPRL13}, and several
pathways to molecule formation imply 
interesting avenues for cold controlled chemical reactions.  

\acknowledgments

We would like to thank Tommaso Calarco for bringing our attention to
this problem and Ren\'e Gerritsma for useful discussions.  
MT and RM thank the Foundation for Polish Science for support within
the START and MISTRZ programs, respectively; RM was supported by the Polish
Ministry of Science and Education through the project N N204 215539.

\end{document}